\def\draftform{DRAFT}
\def\prlform{PRL}
\def\form{PRL}
\ifx\draftform\form
\documentclass[preprint,showpacs,preprintnumbers,superscriptaddress,amsmath,amssymb]{revtex4}
\else
\documentclass[prl,twocolumn,floatfix,showpacs,preprintnumbers,superscriptaddress,amsmath,amssymb]{revtex4}
\fi
\def\figwidth{\ifx\draftform\form 0.8\else0.45\fi}

\usepackage{graphicx}
\usepackage{umlaut}
\usepackage{dcolumn} 
\usepackage{hyperref}
\usepackage{epsfig}
\newcommand{\fig}[1]{Figure~(\ref{fig:#1})}

\newcommand{\figsat}{
\begin{figure}[t]
\centerline{\epsfig{file=Fig1/figure1.eps,width=\figwidth\textwidth}}
\caption{
Decay exponents $\eta_x$ (circles) and $\eta_z$ (squares) for $\Delta
\le 1$ obtained with the DMRG for L=80 in comparison with analytic
results for the thermodynamic limit. OBC and PBC denote open and
periodic boundary conditions, respectively. The OBC and PBC results
for $\eta_x$ coincide on the scale of the figure.  Inset: $A_z=\langle
S^z_0 S^z_{L/2} \rangle$ as a function of $\Delta \ge 1$ (PBC) in
comparison with analytic results. For $\Delta < 1.2$ the length scale
for saturation is larger than the system. }
\label{fig:sat}
\end{figure}
}

\newcommand{\figdata}{
\begin{figure}[t]
\centerline{\epsfig{file=Fig2/fig2-ww.eps,width=\figwidth\textwidth}}
\caption{
Doubly logarithmic plots of replica averaged (a) $\overline{\langle
S_0^+S^-_r \rangle}$ and (b) $\overline{\langle S_0^zS_r^z\rangle}$ for
$\Delta = 0.5$ and various disorder strengths $W$. z-correlation
functions decay algebraically, the degree of correlation increases
with disorder strength, while x-correlations decay algebraically 
for $W < 1$, but exponentially for $W > 1$.}
\label{fig:data}
\end{figure}
\begin{figure}[t]
\centerline{\epsfig{file=Fig3/fig3.eps,width=\figwidth\textwidth}}
\caption{
Logarithmic plots of replica averaged correlation functions in the
Ising regime ($\Delta = 1.5$) 
for various disorder strengths $W$. 
x-correlations decay algebraically for $W < 1$ and exponentially for
$W > 1$, z-correlation functions saturate at a finite value that increases
with the disorder strength (inset) }
\label{fig:ising}
\end{figure}
}

\newcommand{\figresult}{
\begin{figure}[!t]
\ifx\draftform\form
\centerline{\epsfig{file=Fig4/fig4a.eps,width=0.6\textwidth}}
\vspace*{0.6cm}
\centerline{\epsfig{file=Fig4/fig4b.eps,width=0.6\textwidth}}
\else
\centerline{\epsfig{file=Fig4/fig4a.eps,width=\figwidth\textwidth}}
\vspace*{0.6em}
\centerline{\epsfig{file=Fig4/fig4b.eps,width=\figwidth\textwidth}}
\fi

\caption{(a) Exponents $\eta$ for the algebraic decay  of the $x$ and
$z$ correlation functions in the critical regime $\Delta < 1$. In
contrast to the predictions of the RSRG the exponents change
continuously and slowly from their (finite system) values with no
disorder. Circles, diamonds and squares denote $\Delta=0.2, 0.5,$ and
0.8, respectively.  Open symbols correspond to $\eta_z$ and filled
symbols to $\eta_x$. (b) Inverse correlation length for the decay of
the $x$ correlation functions for various $\Delta$ as a function of
$W$. For $W > 1$ all data are consistent with the results for the
exactly solvable model (heavy line). }
\label{fig:result}
\end{figure}
}

\begin{document}
\bibliographystyle{apsrev}
\title{Disorder Induced Quantum Phase Transition in 
Random-Exchange Spin-1/2 Chains}

\author {K. Hamacher} 
\affiliation{Forschungszentrum Karlsruhe, Institut f\"ur Nanotechnologie,
76021 Karlsruhe, Germany}

\author {J. Stolze} 
\affiliation{Institut für Physik, Universität Dortmund, 44221
Dortmund, Germany}

\author {W. Wenzel} 
\affiliation{Forschungszentrum Karlsruhe, Institut f\"ur Nanotechnologie,
76021 Karlsruhe, Germany}

\date{October 24, 2001}

\begin{abstract}
  We investigate the effect of quenched bond-disorder on the
  anisotropic spin-1/2 (XXZ) chain as a model for disorder induced
  quantum phase transitions. We find non-universal behavior of the
  average correlation functions for weak disorder, followed by a
  quantum phase transition into a strongly disordered phase with only
  short-range xy-correlations. We find no evidence for the universal
  strong-disorder fixed point predicted by the real-space
  renormalization group, suggesting a qualitatively different view of
  the relationship between quantum fluctuations and disorder.
\end{abstract}


\maketitle

The existence and nature of quantum phase transitions (QPTs)
\cite{Sac99,SBV99,SY97} has in recent years emerged as one of the
most interesting aspects of low-dimensional quantum systems.  QPTs
arise from the subtle interplay between short-range interactions on
one hand and quantum fluctuations on the other~\cite{LR85}. Since the
latter are particularly strong in one dimension, quantum spin chains
have emerged as a generic model to investigate
QPTs~\cite{CFJxx98,SBCA01,Zvy00,CLI01}. The additional
presence of disorder has profound effects on the properties of low
dimensional systems~\cite{Bha98,USTxx99} as it competes with the
subtle effects of quantum fluctuations. Its effect on QPTs has been
the subject of recent intense and controversial
discussion~\cite{Zvy00,CLI01,SBC00,HRD93,HG98,UM00,KTHxx00,KD01}.

In one-dimension the strong-disorder renormalization
(SDRG)~\cite{MDH79,DM80} group offers potentially exact results for a
variety of models. Of particular interest is the prediction of a
universal infinite randomness fixed point (IRFP) for disordered spin
chains. In many systems SDRG studies suggest the relevance a {\it
random-singlet} (RS) phase\cite{DF92,FFF} as the ground state for
fairly general disorder. In this RS phase the average spin
correlations are predicted to obey a {\em universal isotropic
power-law decay},
 $ |\overline{\langle S_i^{\alpha} S_j^{\alpha} \rangle}|\sim |i-j|^{-2}
  \  (\alpha=x,z)$, 
where the overbar denotes a configurational average over many random
chains (replicas). The Luttinger (or spin-liquid) continuum of
critical ground states of the ordered chain is thus
predicted to collapse to a single point.  Numerical results consistent
with the RS picture were reported for relatively short ($N \leq 18$)
XXZ chains \cite{HRD93} and also for long XX chains ($\Delta=0$)
\cite{HG98} with couplings uniformly distributed in $[0,1]$.
Recently, some studies suggest the relevance of such a fixed
point for the q-state quantum clock model and the quantum Ashkin
Teller model~\cite{CLI01}, while others dispute its existence, at least
for $S > 1/2$~\cite{SBCA01}.  

Recent numerical advances, in particular the development of the
density matrix renormalization group~\cite{PWH99}, now offer a
framework to investigate the relevance of the IRFP to realistic
one-dimensional spin systems~\cite{NLL96}.  In this Letter we
investigate the influence of exchange disorder on the anisotropic spin
1/2 Heisenberg chain (XXZ model), one of the best-known model systems
for QPTs in one dimension. We find a qualitatively different scenario
for the interplay of quantum fluctuations and disorder. Our results
indicate that the spin correlations do {\em not} obey the universal
parameter independent decay law suggested by the RS picture. Instead
we find a disorder-induced QPT for finite disorder strength, whose
nature can be illustrated by an exactly solvable model.

In this paper we present results of a density matrix renormalization
group study of XXZ chains with randomness in the transverse
nearest-neighbor coupling,
\begin{equation}
\label{eq:1}
H=J \sum_{i=1}^{N-1} \left[  \lambda_i (S_i^x S_{i+1}^x + S_i^y S_{i+1}^y) +
    \Delta S_i^z S_{i+1}^z \right],
\end{equation}
($J > 0$) where the anisotropy parameter $\Delta \ge 0$ controls the
relative strength of the quantum fluctuations.  In the homogeneous
system ($\lambda_i \equiv 1$) the ground state of (\ref{eq:1}) shows
long-range order for $\Delta > 1$ (Ising regime) , whereas for $\Delta
\leq 1$ (critical regime) the spin correlations decay algebraically to
zero as
$  |\langle S_i^{\alpha} S_j^{\alpha} \rangle | \sim
|i-j|^{-\eta_{\alpha}} $ 
with non-universal decay exponents \cite{LP75}
\begin{equation}
  \label{eq:3}
  \eta_x=\eta_z^{-1}=1-\pi^{-1} \arccos \Delta.
\end{equation}

The introduction of quenched randomness brings about subtle changes in
the ground-state. For reasons of better numerical control of
replica averages we used a bounded probability density
$  p(\lambda)= \frac{1}{2 W} \Theta(W - |\lambda -1|). $
\ifx\prlform\form \figsat\fi

We investigated chains of length up to $L = 80$ with a
finite-size DMRG algorithm and ensured that the ground state and
correlations $\langle S_i^{\alpha} S_j^{\alpha} \rangle$ could be
determined with consistent accuracy for arbitrary choices of
$\lambda_i$. 
We noted that for long chains the standard DMRG procedure tends to
spontaneously break the local $S_z$ reversal symmetry.  We ensured
that the resulting data for the correlation functions were
nevertheless correct by comparing with explicitly symmetry-adapted
high-accuracy DMRG calculations for a number of randomly selected
replicas $\lambda_i$. We verified the accuracy of the correlation
functions and energies by comparing with exact data for XX ($\Delta =
0)$ chains which can be mapped to non-interacting lattice fermions.
For critical systems (see Fig. \ref{fig:sat}) we investigated finite
size effects on the estimates of the decay exponents $\eta_x$ and
$\eta_z$ (\ref{eq:3}) for $0\le \Delta\le 1$. For ordered chains with
periodic boundary conditions DMRG results for $\langle S_0^z S_{L/2}^z
\rangle$ were in good agreement with
\figdata
the known long-range order parameter for $\Delta \ge 1$ \cite{Bax73}
until the chain length became significantly shorter than the
correlation (or saturation) length \cite{JKM73} of the system (see
inset of Fig.\ref{fig:sat} ). These data suggest that chains of length
L=80 are sufficient to determine the long-range behavior of the
correlation functions to better accuracy than could conceivably be
obtained in the replica average.

We then performed DMRG calculations for a large number of replicas
each for various values of $\Delta$ and $W$.
The number of replicas computed varied from 250 for small $W$,
where replica expectation values fluctuate little, to more than 1000
for large $W$. We have not
gathered replica averaged data for the XX-model  by DMRG as the
correlation functions for a small set of replicas showed perfect
numerical agreement between DMRG and results from exact
diagonalization for chain lengths of up to $L=160$. 

Replica-averaged correlation functions in the critial and Ising regime
are shown in \fig{data} and (3) respectively. The data demonstrates
qualitatively different behavior for small and large values of the
disorder $W$. For $\Delta \le 1$ and $W \le 1$ both $x$ and $z$
correlation functions decay algebraically with exponents that depend
on $\Delta$ and $W$. Fitted to additional exponential components and
finite offsets $|\overline{\langle S_i^{\alpha} S_j^{\alpha} \rangle}|
= A + |i-j|^{-\eta_{\alpha}} \exp(-\gamma |i-j|)$ the data show
negligible inverse correlation lengths $\gamma$ and offsets $A$ for
small $W$ and $\Delta$. The decay of the $x$ correlation accelerates
with growing disorder $W$ whereas that of the $z$ correlation
decelerates. The values of the decay exponents $\eta_x$ and $\eta_z$
as a function of $W$ are shown in \fig{result}(a), indicating a
continuous, but small change of both exponents from their values for
ordered chains $W=0$.

For $W > 1$ the x-correlation functions decay exponentially in both
the Ising and the critical regime. The inverse correlation length of
the $x$ correlation function is shown in \fig{result}(b). The data is
consistent with a crossover to exponential decay at $W = 1$ with
significant finite size effects for $W>0.8$, in particular in the
Ising regime. In the Ising phase ($\Delta > 1$) the $z$ correlation
functions continue to decay to a finite value for large separations
(see \fig{ising} (inset)).

\figresult

The nature of the transition at $W=1$ is explained by a simple exactly
solvable model, defined by the bimodal distribution
\begin{equation}
  \label{eq:7}
  p(\lambda) = p \delta(\lambda+1) + (1-p) \delta(\lambda-1)
\end{equation}
in the general Hamiltonian (\ref{eq:1}) with $\Delta <1$. The ground
state spin correlations of this model are related to the known
correlations \cite{LP75} of the homogeneous chain $(p=0)$ by a simple
gauge transformation \cite{SRM_un,FSLTN94}. Consider a single replica,
i.e. one configuration of $\lambda_i$s drawn from the distribution
(\ref{eq:7}) for an open chain. By a suitable product $U$ of $\pi$
rotations $\exp (2 i \pi S_i^z)$ about the local $S_z$ spin axes the
disorder may be gauged away, i.e. $\tilde{H}=UHU^{\dag}$ describes a
configuration with $\lambda_i \equiv 1$. As the $z$ spin components
are invariant under $U$ the $z$ correlations of the disordered system
are {\em identical} to those of the homogeneous system. In contrast, a
product of two $x$ spin components aquires a string of random signs:
\begin{equation}
  \label{eq:8}
  US_i^xS_{i+r}^x U^{\dag} = S_i^xS_{i+r}^x \prod_{l=i}^{i+r-1} \lambda_l.
\end{equation}
The disorder average of  (\ref{eq:8}) simply yields the $x$
correlation of the homogeneous system, multiplied by $(1-2p)^r$. The
decay of the $z$ correlation thus remains algebraic, whereas the $x$
correlation function is modified by an {\em exponentially decaying}
factor, with a correlation length diverging with critical exponent
equal to unity at the two quantum critical points $p_c=0,1$:
$  \xi_x \sim \frac{1}{2|p-p_c|}.$
Applied to the probability density used in the DMRG calculations, this
argument yields $p=(W-1)/(2W)$ and $\xi^{-1}=\ln W$ for $W>1$
(heavy solid line in Fig. \ref{fig:result}(b)). For $\Delta=0$ data
for larger systems are available \cite{SRM_un} and the crossover from
$\xi^{-1}=0$ to $\xi^{-1}=\ln W$ is visible more clearly.

In the limiting cases that are accessible by alternate techniques, our
results are in good agreement with both exact data for $W=0$ (Fig.
\ref{fig:sat}) and numerical diagonalization results~\cite{SRM_un} for
long ($N \leq 256$) XX chains (see also~\cite{HG98}). The latter also
display clear deviations from the IRFP behavior predicted by the
SDRG. For the $z$ correlation an $r^{-2}$ decay with the
corresponding finite-size scaling behavior~\cite{KMS94} remains
perfectly intact from $W=0$ up to $W =2.$ The static $z$ structure
factor is linear in the wave vector $q$ and independent of $W$. In
contrast, the $x$ correlation does not show finite-size scaling and
the static $x$ structure factor is neither linear nor
$W$-independent. The $x$ correlation decays progressively faster as
$W$ grows. The data may be fitted to a power law as long as $W<1$, but
with an exponent significantly smaller than the value of 2 predicted
for the RS phase. For $W>1$ the decay is exponential, with an inverse
correlation length proportional to the fraction of negative $\lambda$s
(as in the exactly solvable model above).

Combined, these results suggest a qualitative revision of the
influence disorder is thought to have on quantum spin chains. No signs
of attraction to the IRFP (with universal and isotropic algebraic
decay of the spin correlation functions) predicted by the SDRG were
observed in our study. Instead we observe a disorder-driven quantum
phase transition at $W=1$ for $0 \leq \Delta \leq 1$, from a spin
liquid phase with algebraically decaying correlations (with
non-universal exponents) at $W < 1$ to a different phase with
exponential decay of the $x$ correlations. These observations suggest
to critically reexamine the applicability of the SDRG to this system
and to investigate the possible existence of a crossover length-scale
beyond which the IRFP emerges as relevant~\cite{SK_priv}. We hope that
research on thermodynamic properties and spin correlations, in
particular of the numerically much more accessible XX chains, combined
with the investigation of pre-asymptotic behavior in the the SDRG on
the other hand will help resolve this issue.

The mechanism for the destruction of the critical phase by disorder
elucidated in this study may have interesting implications for
experiments. For nonbounded disorder, where fluctuating signs of the
couplings are present with varying probability for any $W$, one may
anticipate the existence of a crossover length scale where algebraic
decay crosses into exponential decay as a function of system size that
may be observable by studying long, but finite chains. DMRG studies
for Gaussian disorder, for which replica averages are much more
difficult to converge, are presently under way to explore this
scenario. The results of the present study are directly applicable to
systems where a crystal field or easy plane generates a natural
anisotropy. Extrapolating from both limits, they suggest the continued
existence of critical behaviour for weak isotropic disorder in the
isotropic model, a scenario that we will investigate more thoroughly
in the future.

JS acknowledges helpful discussions with S. Kehrein (Universit\"at
Augsburg), KH acknowledges the support of the Stipendienfonds der 
Chemischen Industrie, the BMBF and of the Studienstiftung des dt. Volkes and WW
the computational resources of the von Neumann Institute for
Computing.

\bibstyle{unsrt}
\bibliography{random_chain}

\ifx\draftform\form \figsat\fi

\end{document}